\documentclass[twocolumn,showpacs,preprintnumbers,amsmath,amssymb,epsfig]{revtex4}
\usepackage{epsfig}
\begin{document}

\def\omunit{{(km s$^{-1}$)/kpc}}
\def\gtrsim{ \lower .75ex \hbox{$\sim$} \llap{\raise .27ex \hbox{$>$}} }
\def\lesssim{ \lower .75ex \hbox{$\sim$} \llap{\raise .27ex \hbox{$<$}} }

\def\sun{\odot}
\def\A{{\sf A}}
\def\B{{\sf B}}
\def\Ms{\mathrm{M}_\sun}
\def\vk{v_\mathrm{kick}}
\def\ee{{\it e}}
\def\vcm{v_{CM}}
\def\vcm0{v_{CM0}}
\newcommand\arcsec{\mbox{$^{\prime\prime}$}}%

\def\figurestyle#1{\sf \small \baselineskip 14pt #1}


 \title{A remark on ``On the Formation and Progenitor of PSR
 J0737-3039: New Constraints on the Supernova Explosion Forming
 Pulsar B" by Willems et al., \cite{ref6}
 }
 \author{Tsvi Piran \& Nir J. Shaviv}
\email{shaviv@phys.huji.ac.il, piran@phys.huji.ac.il}
 \affiliation{
 Racah Institute of Physics, Hebrew University, Jerusalem 91904,
Israel }
\begin{abstract}

Shortly after the discovery of the binary system PSR J0737-3039
\cite{ref1,ref2} we \cite{PSastroph,PSprl} predicted that it has a
small proper motion and suggested that this implies that the
progenitor mass of  the younger pulsar (\B) must be around
1.45~$\Ms$. This is in contradiction to standard evolutionary
scenarios that suggest a He-star progenitor  with $M >
2.1-2.3~\Ms$ \cite{ref3,ref4} and in sharp contrast to a velocity
of $\sim 100 km/s$ predicted by these models \cite{ref4}. It
requires a new previously unseen type of gravitational collapse.
Our (generally ignored) prediction of a low peculiar motion was
confirmed later by pulsar timing observations \cite{ref4a} that
found $v_\bot < 30$ km/s. Recently Willems et al. \cite{ref6} put
forward an elaborate simulation suggesting that in spite of these
observations, the most likely natal kick is large (50-170 km/s)
and the corresponding most likely progenitor mass is large as
well. We refute this claim and show that Willems et al.
\cite{ref6} implicitly assume the result that they obtain.

\end{abstract}

%






\maketitle


Folllowing the discovery of the binary system PSR J0737-3039
\cite{ref1,ref2} we \cite{PSastroph,PSprl} realized that its
location, $\sim 50$~pc from the Galactic plane, implies that the
velocity of the system is very small. This, in turn, poses strong
limits on the natal kick velocity of the system and on the
progenitor mass of the younger pulsar (\B). The kinematically
favored progenitor mass is around 1.45~$\Ms$. This low mass
requires a new previously unseen gravitational collapse mode. Our
prediction that the system has a small proper motion was in
contradiction to the predictions of the standard model
\cite{ref4,ref3} that assumed that the progenitor was a He-star with
$M > 2.1-2.3~\Ms$. This  would have implied a typical proper
motion of more than 100~km/s.

Our prediction of a very small proper motion was indeed verified
when pulsar timing \cite{ref4a} set an upper limit of 30~km/s on
the proper motion of this system \footnote{This discovery came
after erroneous claims of much larger peculiar velocity
\cite{ref11new}.}. In spite of that, Willems et al. \cite{ref6} presented
elaborate simulations suggesting that the most likely
natal kick is large (50-170 km/s) as is the most likely progenitor
mass. They reconcile these suggestions with the limit on
the proper motion by assuming that the system is moving with a
very large velocity almost exactly towards us. Willems et al.
\cite{ref6} insinuate that the difference between our conclusions
and theirs arises because of the ``higher accuracy" of their
calculations that involve a three dimensional tracking of the
galactic motion of the system whereas we \cite{PSprl} consider
only the vertical motion. We refute this claim and show that
Willems et al. \cite{ref6} implicitly assume the result that they
obtain.

We begin with a brief simplified outline of our reasoning. For
brevity we skip important but non essential details \cite{PSprl}.
If a significant mass is ejected during the second supernova
collapse the binary system will be disrupted unless the second
neutron star gets a large (comparable to its orbital velocity)
natal kick in the right direction. Getting the post SN system into
an almost circular orbit (as observed in J0737-3039) requires
additionally a specially tuned direction. The kick given to the
newborn pulsar will also  be reflected by a large CM velocity,
$\vcm0$, of the binary system, which will be of order of a third
of the orbital motion of the resulting system (this orbital
velocity is $\sim 600$km/s for PSR J0737-3039). Determination of
$\vcm0$, the CM velocity right after the second supernova can
therefore set an upper limit on the ejected mass and from there on
the progenitor's mass.

It is impossible, of course, to measure $\vcm0$. But if we can
estimate the current CM velocity, $v_{CM}$  we would have a first
approximation for $\vcm0$. One can expect that in a typical
galactic motion the magnitude of the planar components of peculiar
motion will not change much \footnote{The vertical motion is easy
to trace.}. To estimate $v_{CM}$ we used, in 2004, the location of
the system $z_{obs}\approx 50$pc from the plane of the Galaxy.
Stars move in a periodic motion in the vertical direction. For
small vertical oscillations, the potential of the Galaxy is
harmonic with a vertical orbital period, $P_z \approx 50$~Myr. The
typical velocity for an object at $z_{obs}$ is $v_z \approx 2\pi
z_{obs}/P_z$. $z_{obs} \approx 50$pc implies that the expectation
value of the vertical velocity is of the order of 6km/s. This
reasoning has lead us to predict that both $v_\bot$, the proper
velocity  of the system that was not known at that time, and
$v_{CM}$  will be small.

By now, this stage is not necessary. We know  that $v_\bot<30$km/s
\cite{ref4a}. This is a much stronger result than what we had in
2004. First the statistical uncertainty involved with  the
possibility that the system is just accidentally in the galactic
plane disappears. Moreover, the vertical motion constrained just
one component of the velocity whereas the present limit of
$v_\bot$ constrains two.

However, we still need to know the full magnitude of $v_{CM}$.
Using Bayes' theorem we write:
\begin{equation}
P(v_{CM} | v_\bot<30{\rm km/s}) \propto P( v_\bot<30{\rm km/s}|
v_{CM} ) P(v_{CM}) . \label{bayse}
\end{equation}
Simple geometric arguments allow us to estimate the first factor
on the rhs:
\begin{equation}
P(v_\bot<30{\rm km/s}| v_{CM} ) = 1-\cos \left ({30{\rm ~km/s}\over
v_{CM}}\right ) \label{prob} .
\end{equation}
If we are careful not to  inflict our theoretical bias on the
resulting distribution we should not have any prior on the
probability distribution of $v_{CM}$. In this case, as expected a
small observed $v_\bot$ implies a small $v_{CM}$.  We find $v_{CM}
< 150, 100$~km/s at 1\% and 5\% levels respectively. The most
likely value is of course much smaller. This basically leads to
our \cite{PSastroph,PSprl} conclusion that $v_{CM0}$ was small
from which it follows that the corresponding natal kick was small
and finally that the progenitor's mass must have been small as
well. In fact our best bet progenitor mass 1.45~$\Ms$ corresponds
to ejection of 0.2~$\Ms$ (the expected minimal mass loss in form of
neutrinos). Both the observed eccentricity and the limit on
$v_\bot$ are compatible with such a collapse with no kick velocity
at all.

In order to estimate the effects of three dimensional Galactic
motion \footnote{While the vertical motion is very important we
expect that in a typical case the planar motion won't make a
significant change on the magnitude of the velocity.} Willems et
al. \cite{ref6} integrate backwards the orbital motion of the
system from its present place towards its birth place. As such
system are typically born within the Galactic plane they identify
possible birth places as the intersection of the orbit with the
galactic plane. Once  the system's velocity in its birth place
$v_{CM0}$ is known, they proceed to estimate the natal kick and
the progenitor's mass. Repeating many times they obtain a
distribution of velocities and masses.

This procedure, which seems reasonable at first glance, requires a
knowledge of the current velocity and in particular of the
 present radial component of the CM velocity. However, this is
exactly what we are trying to infer from the data. Willems et al.
\cite{ref6}  realize this \footnote{Following an earlier exchange
between us and a subgroup of the authors of \cite{ref6}.} and
consider several distributions of the radial velocity. These
distribution are either ad hoc (a uniform velocity distribution
from -1500 km/s to +1500 km/s) or derived from a population
synthesis model (Gaussian with  a velocity dispersion 60 km/s - 200
km/s).

In the language of our analysis these {\it assumed} distributions
amount to  priors on $P(v_{CM})$ that appears in Eq. \ref{bayse}.
Clearly with a suitable prior one can get any result one wants for
$P(v_{CM} | v_\bot<30{\rm ~km/s})$ regardless of how small is the
measured proper velocity and if $v_{CM}$ is large both the kick
velocity and the progenitor's mass are large. Thus, Willems et al.
\cite{ref6} simply {\it assume} (via this prior) the result that
they obtain.

There is not much to say about the assumption of an ad hoc uniform
velocity distribution. It yields typically large velocities (less
than a 10\% chance of a velocity less than 100 km/s) forcing the
distribution to favor large natal kicks and large progenitor
masses. The distribution derived from the population synthesis is
more subtle as it presumably follows a valid theoretical
justification. However, the population synthesis calculations are
based on the current standard model for binary NS formation, and are hence implicitly biased towards large progenitor masses. Namely, all that
the authors have done was to {\it assume} a large progenitor mass,
simulate forwards in time to find that it implies a large radial
velocity, that must be well aligned with the line of sight towards the observer. They then feed this velocity as input and integrate
backwards in time and find, to ones great surprise, a large
progenitors mass. At least the calculations are consistent.

\def\mnras{Mon.\ Not.\ Roy.\ Astr.\ Soc.}
\def\apj{Ap.\ J.}
\def\apjl{Ap.\ J.}


\end{document}